\documentclass[12pt,preprint]{aastex}
\usepackage{psfig,lscape}
\def\lap{\lower.5ex\hbox{$\; \buildrel < \over \sim \;$}}
\def\gap{\lower.5ex\hbox{$\; \buildrel > \over \sim \;$}}

\def\ergcm2s{${\rm erg\ cm^{-2}\ s^{-1}}$}

\def\ergscm2s{${\rm erg\ cm^{-2}\  s^{-1}}$}

\def\cm-2{${\rm cm^{-2}}$}

\begin{document}

\title{{\it Chandra} and {\it Hubble} Study of a New Transient X-ray Source in M31}

\author{Benjamin F. Williams\altaffilmark{1}, Michael R. Garcia\altaffilmark{1}, Jeffrey E. McClintock\altaffilmark{1}, Frank A. Primini\altaffilmark{1}, and Stephen S. Murray\altaffilmark{1}}
\altaffiltext{1}{Harvard-Smithsonian Center for Astrophysics, 60
Garden Street, Cambridge, MA 02138; williams@head.cfa.harvard.edu;
garcia@head.cfa.harvard.edu; jem@head.cfa.harvard.edu; fap@head.cfa.harvard.edu; ssm@head.cfa.harvard.edu}

\begin{abstract}

We present X-ray and optical observations of a new transient X-ray
source in M31 first detected 23-May-2004 at R.A.=00:43:09.940 $\pm
0.65''$, Dec.=41:23:32.49 $\pm 0.66''$.  The X-ray lightcurve shows
two peaks separated by several months, reminiscent of many Galactic
X-ray novae. The location and X-ray spectrum of the source suggest it
is a low mass X-ray binary (LMXB).  Follow-up {\it HST} ACS
observations of the location both during and after the outburst
provide a high-confidence detection of variability for one star within
the X-ray position error ellipse.  This star has $\Delta B \approx 1$
mag, and there is only a $\approx$1\% chance of finding such a
variable in the error ellipse.  We consider this star a good candidate
for the optical counterpart of the X-ray source.  The luminosity of
this candidate provides a prediction for the orbital period of the
system of 2.3$^{+3.7}_{-1.2}$ days.

\end{abstract}

\section{Introduction}

X-ray and optical investigations of bright transient Galactic low-mass
X-ray binaries (LMXBs), known as X-ray novae (XRNe), have shown these
sources to be some of the best stellar-mass black hole candidates
known (\citealp{mcclintock2004}, and references therein).  During
outburst, some of these objects are bright enough in X-rays
($\gap10^{38}$ erg s$^{-1}$) to be easily detected in nearby galaxies,
and each new X-ray transient source found is a potential gem for
studies of black hole and accretion physics.

Over the past several years, we have undertaken a monitoring program
of the nearby spiral galaxy M31 to search for new transient X-ray
sources with {\it Chandra} and {\it HST}.  Through this program, and
the searches done by other groups using {\it XMM-Newton}, dozens of
new transient sources have been found
(e.g. \citealp{williams2004hrc,distefano2004,kong2002acis,trudolyubov2001,osborne2001};
Williams et al. in preparation).

While the discovery of new XRN candidates is exciting, most of the
X-ray transient sources detected in M31 so far have no known optical
counterparts, making their classification and orbital parameters
difficult to constrain.  Our coordinated {\it HST} program has
provided several of the first optical counterpart candidates for these
transient X-ray sources, likely to be XRNe
\citep{williams2005bh1,williams2005bh2,williams2005bh4,williams2004hrc}.
Knowledge of the optical counterparts begins the process of
constraining the orbital periods of the binaries.  As more
X-ray/optical systems are found, we will be able to study the orbital
period distribution for XRNe in M31.  This distribution is one of the
fundamental observable parameters for this class of objects that any
viable model of their evolution must be able to reproduce.

This paper describes our discovery and follow-up study of a new
transient X-ray source in M31 detected by {\it Chandra} in
observations spanning from 23-May-2004 through 04-Oct-2004.  In
addition, we describe our efforts to determine the optical counterpart
of the source using {\it HST}.  Section 2 describes the data obtained
and the methods used to process them.  Section 3 provides our results.
Section 4 discusses the implications of the results, including a
prediction of the orbital period of the LMXB, and \S 5 gives our
conclusions.

\section{Data}

\subsection{X-ray}

We obtained six observations with the {\it Chandra} ACIS-I between
December 2003 and December 2004 relevant to this study.  The
observation identification numbers, dates, pointings, roll angles, and
exposure times of these observations are given in Table~\ref{xobs}.
Although the observations were taken for 5 ks each, the effective
exposure was reduced by $\sim$20\% because the data were taken in
``alternating readout mode'' in order to avoid pileup for any bright
transient source.

These observations were all reduced using the software package CIAO
v3.1 with the CALDB v2.28.  We created exposure maps for the images
using the task {\it
merge\_all},\footnote{http://cxc.harvard.edu/ciao/ahelp/merge\_all.html}
and we found and measured positions and fluxes of the sources in the images using
the CIAO task {\it
wavdetect}.\footnote{http://cxc.harvard.edu/ciao3.0/download/doc/detect\_html\_manual/Manual.html}
Each data set detected sources down to (0.3--10 keV) fluxes of
$\sim$6$\times$10$^{-6}$ photons cm$^{-2}$ s$^{-1}$ or 0.3--10 keV
luminosities of $\sim$10$^{36}$ erg s$^{-1}$ for a typical X-ray
binary system in M31.

We aligned the coordinate system of the X-ray images with the optical
images of the Local Group Survey (LGS; \citealp{massey2001}).  The
positions of X-ray sources with known globular cluster counterparts
were aligned with the globular cluster centers in the LGS images using
the IRAF\footnote{IRAF is distributed by the National Optical
Astronomy Observatory, which is operated by the Association of
Universities for Research in Astronomy, Inc., under cooperative
agreement with the National Science Foundation.} task {\it ccmap}.
The errors of this alignment were typically $\sim$0.1$''$.  The
precise alignment errors for the observations used to determine the
X-ray error circle are given in Table~\ref{xpos}.  

We checked the positions of all of the sources we detected in each
observation with those of previously cataloged X-ray sources in M31
and with the {\it Simbad}\footnote{http://simbad.u-strasbg.fr/}
database.  Herein we focus on one bright source that did not appear in
any previous catalogs.  This position also had no cataloged source in
{\it Simbad} within 10$''$.  We name this source
CXOM31~J004309.9+412332, according to the IAU-approved naming
convention described in \citet{kong2002acis} and r3-127 according to
the short naming scheme described in \citet{williams2004hrc}.  This
position is s 4.8$'$ east and 7.4$'$ north of the M31 nucleus, on the
outskirts of the M31 bulge.

We measured the position errors for r3-127 using the IRAF task {\it
imcentroid}.  These errors were affected by the size of the {\it
Chandra} point spread function 9$'$ off-axis and were $\sim$1$''$ for
the brightest detections.  The precise values for these errors are
provided in Table~\ref{xpos}.  Upper-limits to the X-ray flux at this
position in observations 4680 and 4723 were measured by determining
the flux necessary to produce a detection 3-$\sigma$ above the
background flux.

Finally, we extracted the X-ray spectrum of r3-127 from observations
4682 and 4721 using the CIAO task {\it
psextract},\footnote{http://cxc.harvard.edu/ciao/ahelp/psextract.html}
binning the spectrum in energy so that each bin contained $\gap$10
counts.  We then fit each spectrum using CIAO 3.1/Sherpa, trying both
a power-law model with absorption and a disk blackbody model with
absorption.  The spectrum did not contain enough counts to constrain
the foreground absorption.  We therefore fixed the absorption to the
typical Galactic foreground value (6$\times$10$^{20}$ cm$^{-2}$).
This value allowed good fits for both model types.  Results are given
in Table~\ref{spectab} and discussed in \S~3.

\subsection{Optical}

We obtained two sets of observations of the position of r3-127 using
the {\it HST} ACS.  Each data set was obtained in a single orbit and
was observed through the F435W ($B$ equivalent) filter using the
standard 4-point box dither pattern to allow for optimal spatial
resolution.  Each data set was observed with a total exposure time of
2200 seconds.  The first of these data sets was obtained 02-Nov-2004,
while the X-ray source was likely still active.  The second was
obtained 01-Jan-2005, after the source had faded below our {\it
Chandra} detection limit ($\sim$10$^{36}$ erg s$^{-1}$).

Each data set was combined into a final, high-resolution photometric
  image using the PyRAF\footnote{PyRAF is a product of the Space
  Telescope Science Institute, which is operated by AURA for NASA.}
  task {\it multidrizzle},\footnote{multidrizzle is a product of the
  Space Telescope Science Institute, which is operated by AURA for
  NASA. http://stsdas.stsci.edu/pydrizzle/multidrizzle} which has been
  optimized to process ACS imaging data.  The task removes cosmic ray
  events and geometric distortions, and it drizzles the dithered
  frames together into a final, high-resolution, photometric image.

We aligned the final $HST$ images to the LGS coordinate system with
{\it ccmap} using stars common to both data sets. The stars positions
in both the LGS and ACS images were determined with {\it
imcentroid}. The resulting alignment had rms errors of $\leq$0.03$''$
(less than 1 ACS pixel).  The ACS images, independently aligned with
the LGS coordinate system, are shown in Figure~\ref{ims}.  Resolved
stellar photometry was then performed on the relevant sections of the
final images with DAOPHOT II and ALLSTAR \citep{stetson}.  The count
rates measured from our images were converted to VEGA magnitudes using
the conversion techniques provided in the ACS Data
Handbook\footnote{http://www.stsci.edu/hst/acs/documents/handbooks/DataHandbookv2/ACS\_longdhbcover.html}.

\section{Results}\label{results}

\subsection{X-ray}

Source r3-127 was clearly detected 4 times by our monitoring program.
These observations provided a lightcurve, precise positional
constraints, and X-ray spectral measurements.

The X-ray lightcurve of r3-127 is shown in Figure~\ref{lc}, and the
measured fluxes and hardness ratios are provided in Table~\ref{flux}.
Observations before the first detection allowed reliable upper limits
to the X-ray flux from that position to be measured.  Our final
observation yielded a marginal detection with a signal-to-noise of 2.
We treated this measurement as a detection keeping in mind that if it
was spurious the high end of the error range gives the 3$\sigma$
upper-limit of the flux during the final observation
(2004-December-05).  The lightcurve is complex, including a
double-peak, as has been observed for several Galactic XRNe (see
lightcurves in \citealp{mcclintock2004}).  There was a factor of 3
drop in flux followed by a factor of 4 recovery between the May and
October observations.  The brightest detection was the second peak,
with a flux of (7.4$\pm$0.8) $\times$ 10$^{-5}$ photons cm$^{-2}$
s$^{-1}$.

The source faded by at least a factor of eight in the 55 days after
the second peak.  The 3$\sigma$ flux upper limit for the
2004-December-05 observation (9$\times$10$^{-6}$ ph cm$^{-2}$
s$^{-1}$) yields an $e$-folding decay time after the second peak of
$<$1 month.  If the 2$\sigma$ detection of (6$\pm$3$)\times$10$^{-6}$
ph cm$^{-2}$ s$^{-1}$ in the 2004-December-05 observation is real,
then the $e$-folding decay time was $\sim$22 days.  This decay was
faster than that observed after the first peak, where the flux decayed
by a factor of 2.8 in 55 days, exhibiting an $e$-folding decay time of
$\sim$54 days.

Although r3-127 was 9$'$ off-axis, the {\it Chandra} images provided a
precise source position. We measured the position in the two brightest
detections. Table~\ref{xpos} shows the significant sources of error in
this measurement.  These were the alignment errors, determined by {\it
ccmap}, and the position errors, determined by {\it imcentroid}.  We
added these errors in quadrature for each measurement.  Then we took
the weighted mean R.A. and Dec. to obtain our final position and
position error of the source in the LGS coordinate system, which we
used to plot the 1$\sigma$ error ellipses shown in Figure~\ref{ims}.
The final X-ray position and errors were R.A = 0:43:09.940 $\pm$
0.65$''$ and Dec. = 41:23:32.49 $\pm$ 0.66$''$.  At the distance of
M31 (780 kpc), the semi-major axis of the error ellipse is 2.5 pc.

The two brightest detections contained sufficient counts to perform
fits to the X-ray spectrum of r3-127.  The results of the fits are
given in Table~\ref{spectab}.  They show that, while the spectrum was
well-fitted by both the absorbed disk blackbody and the absorbed
power-law models, it was better fitted by the absorbed disk blackbody
with an absorption-corrected 0.3--7 keV luminosity of
$\sim$1.1$\times$10$^{37}$ erg s$^{-1}$ at the time of the brightest
detection.  The spectrum was soft, and it may have become softer
during the decay.  This softening can be seen in the hardness ratios,
as HR1 declined from 0.6$\pm$0.2 at the first detection to 0.0$\pm$0.4
during the decay.  There is also a hint that the second peak may have
been softer than the first in both the hardness ratios and the
spectral fits (see Tables~\ref{flux} and \ref{spectab}). These soft
spectra are typical of Galactic LMXBs, especially those systems that
contain a black hole primary
(e.g. \citealp{tanaka1995,church2001,mcclintock2004}).

Concisely, the X-ray lightcurve of r3-127 was complex, having at least
two peaks.  The position was measured from the {\it Chandra} images
with 0.7$''$ precision, and the spectral fits provided an estimate of
the peak absorption-corrected 0.3--7 keV luminosity reached on
04-Oct-2004 of $\sim$1.1$\times$10$^{37}$ erg s$^{-1}$.

\subsection{Optical}

\begin{figure}
\centerline{\psfig{file=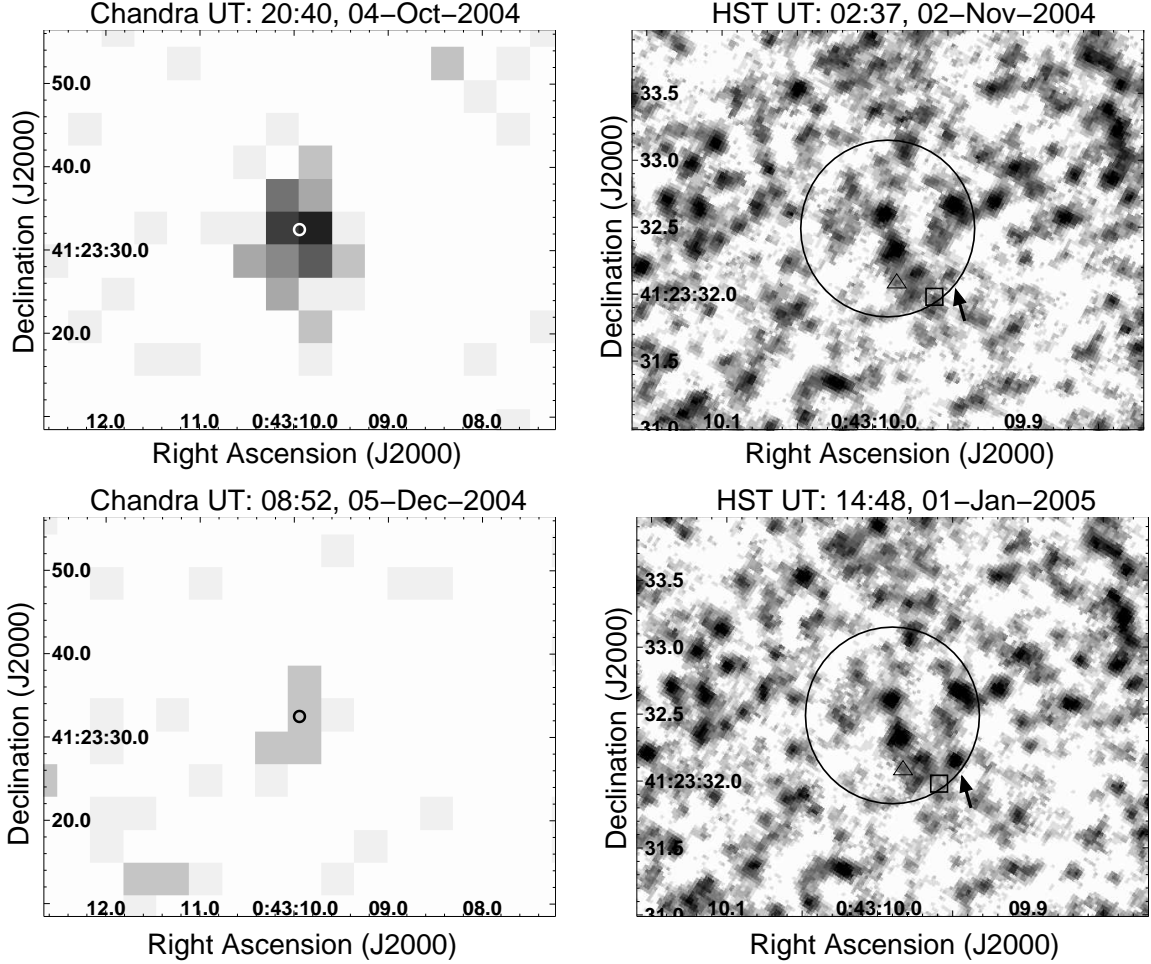,width=6.0in,angle=0}}
\caption{{\it Top left:} The {\it Chandra} image of brightest X-ray
detection of r3-127 is shown with the 1$\sigma$ position error ellipse
overplotted in white.  {\it Top right:} The {\it HST} ACS F435W image
of the position of r3-127 taken less than 1 month after the brightest
X-ray detection.  The black ellipse marks the X-ray position error
(semi-major axis = 0.66$''$, $\sim$2.5 pc), and the arrow marks the
highest amplitude variable star inside the error ellipse.  The black
box marks the only source that showed a statistically significant drop
in brightness between the two epochs not easily attributable to
completeness.  The black triangle marks another faint star that varied
according to our DAOPHOT analysis.  {\it Bottom left:} The {\it
Chandra} image of our final X-ray observation of the position of
r3-127 is shown with the 1$\sigma$ position error ellipse overplotted
in white.  Although there are 10 counts within a 10$''$ aperture
around the location of the source, these were not sufficient to
provide a 3$\sigma$ detection.  These counts provide a signal-to-noise
ratio of 2 for a flux of (6$\pm$3$)\times$10$^{-6}$ ph cm$^{-2}$
s$^{-1}$, corresponding to a luminosity of 1.2$\times$10$^{36}$ erg
s$^{-1}$, showing that the source had faded by at least a factor of
eight in the two months between observations. {\it Bottom right:} The
{\it HST} ACS F435W image of the position of r3-127 taken less than 1
month after the X-ray non-detection.  The ellipse, arrow, box, and
triangle mark the same stars as in the upper-right panel.}
\label{ims}
\end{figure}

\begin{figure}
\centerline{\psfig{file=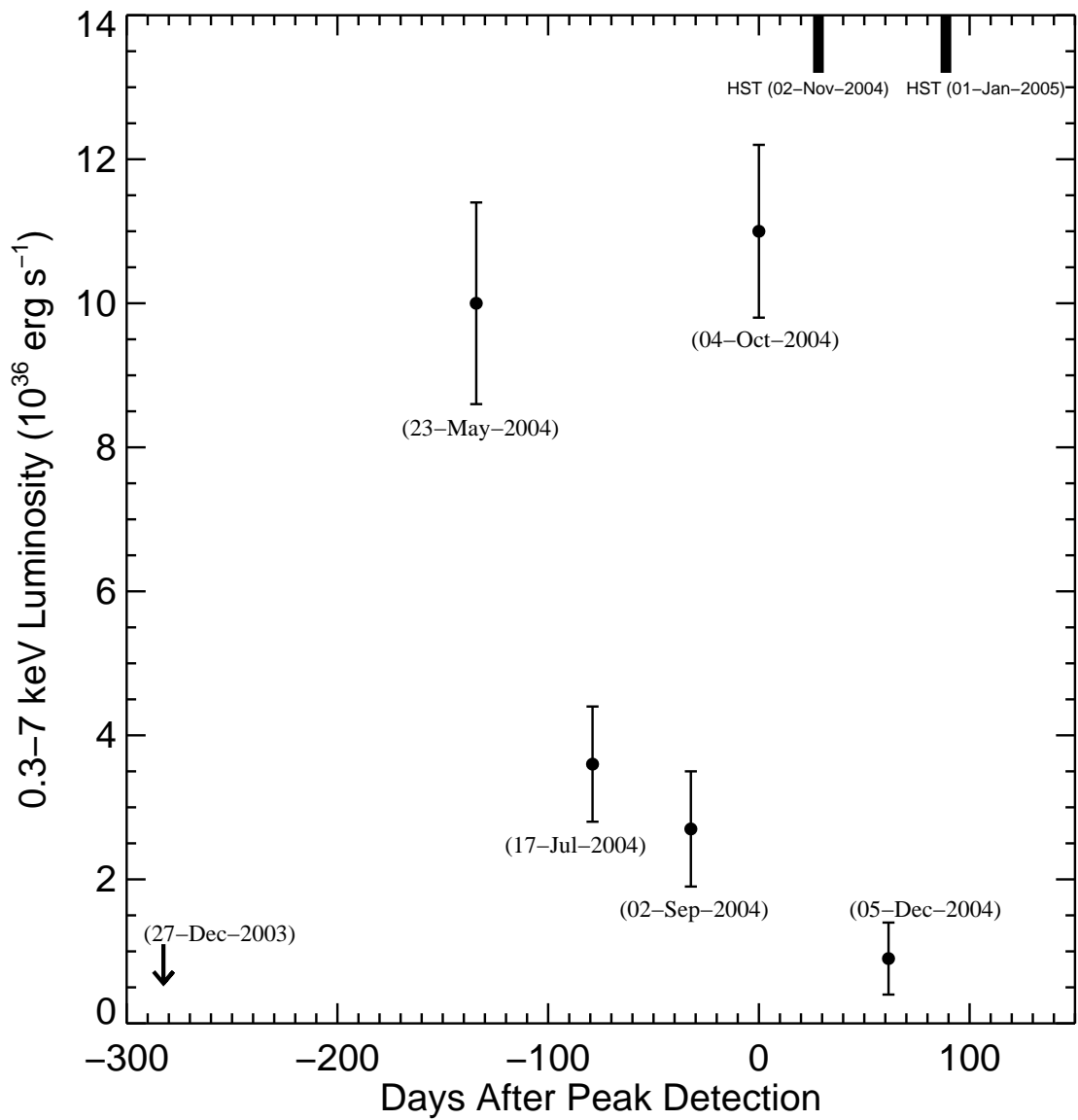,width=6.5in,angle=0}}
\caption{The X-ray lightcurve of r3-127.  The source was active for
$\sim$ 5 months showing two clear peaks in luminosity.  The arrow
marks the upper-limit measured from the observation closest to the
first detection.  Each data point is labeled with its observation
date.  Long, labeled tick marks show the timing of our coordinated
{\it HST} observations.  The final data point is a 2$\sigma$ peak
detection of the source.  If the detection is false, the top of the
error bar on this point is the 3$\sigma$ upper-limit to the luminosity
of the source during our final observation.}
\label{lc}
\end{figure}

The two {\it HST} ACS images of the position of r3-127 are shown in
Figure~\ref{ims} adjacent to the most contemporaneous {\it Chandra}
ACIS~I images available. These data reveal one bright variable source
in the southwest part of the error ellipse.  By analyzing the
completeness of our photometry and carefully examining all potential
variable stars, we were able to justify focusing on this bright
variable as the most likely optical counterpart of r3-127.

\begin{figure}
\centerline{\psfig{file=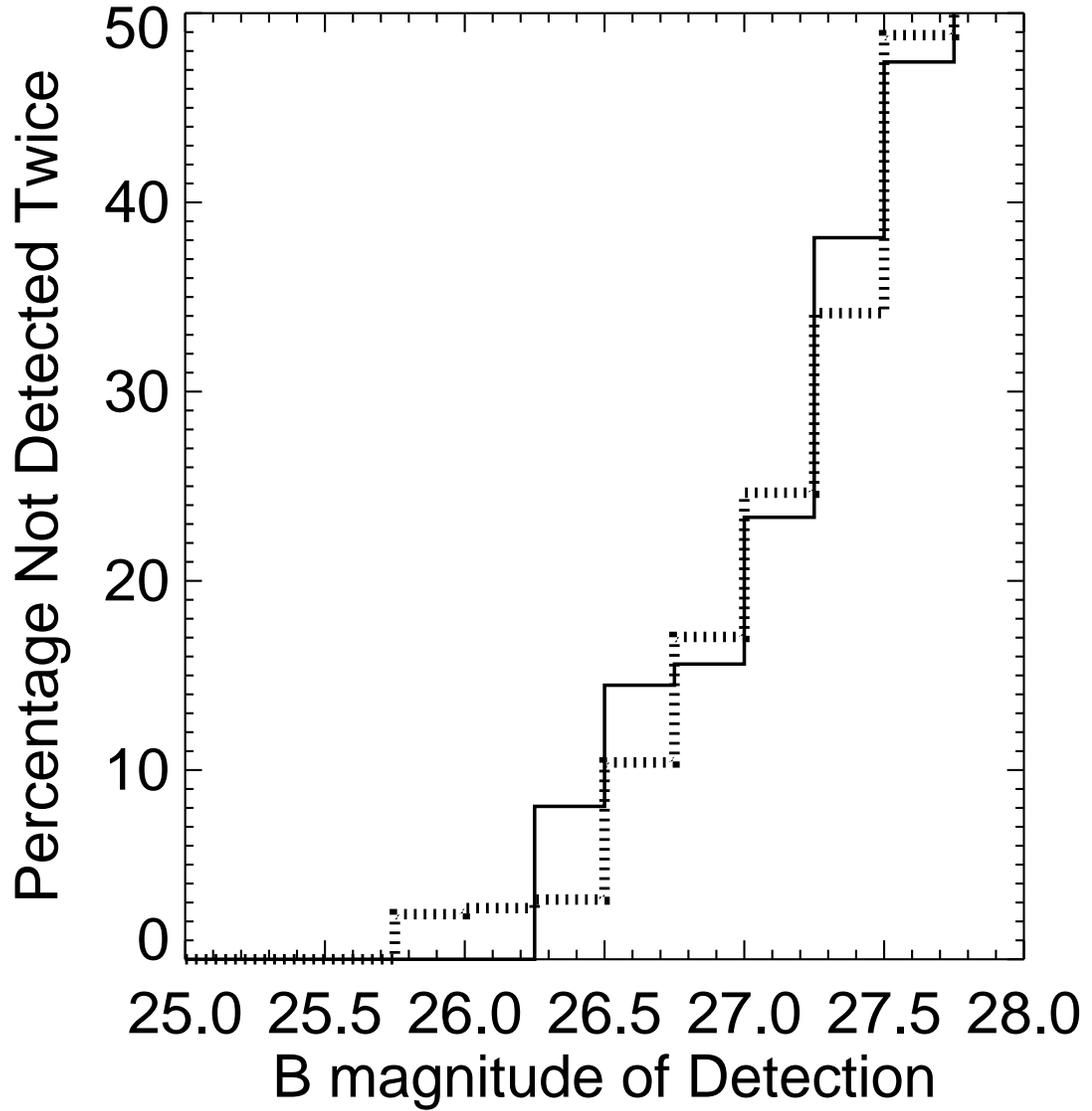,width=6.5in,angle=0}}
\caption{{\it Solid histogram:} The fraction of stars detected by
DAOPHOT and ALLSTAR in the first epoch of ACS data, but not detected
in the second epoch, as a function of the magnitude of the detection.
{\it Dotted histogram:} The fraction of stars detected by DAOPHOT and
ALLSTAR in the second epoch of ACS data, but not detected in the first
epoch, as a function of the magnitude of the detection.}
\label{comp}
\end{figure}

The completeness of our photometry in the two epochs was determined by
comparing the DAOPHOT results from the two epochs of {\it HST} data.
Figure~\ref{comp} shows the results of a completeness analysis of the
ACS data in the region within 3$''$ of the X-ray position of r3-127.
The solid histogram shows the percentage of stars that were detected
in the first epoch and not in the second epoch, as a function of
magnitude.  The dotted histogram shows the percentage of stars that
were detected in the second epoch and not in the first epoch, as a
function of magnitude.  The completeness begins to fall off at
$B\sim26.5$, so that our confidence in the photometry also begins to
decrease at this magnitude.  The completeness falls below 50\% at
$B=27.8$.  We therefore took $B=27.8$ to be our limiting magnitude,
assigning $B>27.8$ as the upper-limit to all non-detections.

A search for variable stars in the error ellipse of r3-127 was then
performed.  There were 10 stars detected inside the error ellipse by
our DAOPHOT analysis of the first epoch that were more than 4$\sigma$
brighter than $B=27.8$ and changed in brightness by $>$4$\sigma$ by
the second epoch.  Seven of the variable candidates were fainter than
$B=26.9$ during the first epoch, and they were not detected in the
second epoch, likely due to completeness.  We therefore removed these
7 stars from the pool of variables and possible counterparts.

Two of the other three candidate variable stars had $B>26.5$ in at
least one epoch.  At these faint magnitudes the photometry, like the
completeness, was likely significantly affected by crowding issues
(see Figure~\ref{comp}).  Such faint stars are more likely to be
blended with stars of similar brightness, which could change the
photometry by $\sim$0.7 mag. 

One of these two variable candidates, marked with the triangles in
Figure~\ref{ims}, was $B=26.80\pm0.09$ in the first epoch and
$B=26.20\pm0.10$ in the second.  It showed only a 4.5$\sigma$ increase
in brightness between epochs with $\Delta B= 0.60\pm0.13$ mag. As the
variability was not highly significant and the photometry likely
affected by crowding, we removed this star from the pool of possible
counterparts.  

The other faint variable candidate, marked with the boxes in
Figure~\ref{ims}, showed a 4.9$\sigma$ drop in brightness between
epochs as measured by DAOPHOT.  The star was $B=26.47\pm0.08$ in the
first epoch and $B=27.09\pm0.09$ in the second, making it the
brightest star that varied in concert with the X-ray flux.  However,
the star only faded by $\Delta B= 0.62\pm0.13$ mag, and its photometry
could be significantly affected by crowding.  We therefore removed it
from the pool of possible counterparts.

The only optical variable candidate left in the r3-127 error ellipse,
marked with the arrow in Figure~\ref{ims}, is clearly seen by visual
inspection of the two images. This candidate was $B=26.27\pm0.07$ in
the first epoch and $B=25.29\pm0.03$ in the second epoch.  This star
exhibited a change in brightness of 0.98$\pm$0.07 mag.  It varied with
13$\sigma$ significance, clearly separating itself from the other
candidate variables, which all showed variations of $<$5$\sigma$.  The
large difference in significance distinguished this variable from the
other candidates, making it our strongest candidate counterpart for
r3-127.

Succinctly, three stars inside the r3-127 error ellipse had
statistically significant ($>$4$\sigma$) brightness changes not easily
attributable to completeness limitations.  Two of these stars were
removed as counterpart candidates of r3-127 because they had
comparable brightness changes only slightly above threshold (4.5--4.9
$\sigma$) and their photometry was likely significantly affected by
crowding.  The only remaining candidate (shown by the arrow in
Figure~\ref{ims}) showed variability of much higher significance than
any other star in the error ellipse.  This candidate is therefore our
strongest counterpart candidate for r3-127 and the only candidate that
will be considered for the remainder of the paper.

\section{Discussion}\label{discussion}

Since most Galactic bright transient X-ray sources are either HMXBs or
LMXBs, it is likely that r3-127 is one of these types of sources.  Our
X-ray and optical data suggest that it is an LMXB.  If r3-127 is an
LMXB and our candidate counterpart is correct, our data can be used to
predict the orbital period of the binary.

The brightest stars in the error ellipse of r3-127 have $B=25.0$.
Assuming a distance modulus of M31 of 24.47 (780 kpc;
\citealp{williams2003}) and typical foreground extinction toward M31
of $A_B = 0.4$, these stars have M$_B\sim0.1$.  This absolute
magnitude is fainter than high-mass O and B stars which are the
typical secondaries of high-mass X-ray binaries (HMXBs).  Therefore
both the faintness of any potential optical counterpart and the X-ray
spectrum (see \S~\ref{results}) suggest that r3-127 is an LMXB.

The candidate counterpart was bright when our X-ray data suggest the
source was faint, which is unexpected but does not rule the star out
as the optical counterpart of r3-127. Such high-amplitude variables at
these magnitudes in 2-epoch ACS photometry of M31 have a density of
$\sim$27 arcmin$^{-2}$ \citep{williams2005var}.  As our error circle
covers only 3.8$\times$10$^{-4}$ arcmin$^2$, there is only a $\sim$1\%
probability of such a variable randomly falling in our error circle.

In addition, the lightcurve of r3-127 (see Figure~\ref{lc}) did not
exhibit a simple monotonic decay over one to several months.  At the
very least, r3-127 exhibited a double-peak reminiscent of XTE~1550-564
\citep{jain2001}. The low sample rate of our lightcurve and the
differing decay times of the peaks highlight the possibility that
r3-127 was exhibiting fast flaring, as has been seen in some Galactic
transient events such as, for example, GRO~1655-40 (see Fig. 1 of
\citealp{remillard1999}) and GRS~1915+105 (see Fig. 1 of
\citealp{rau2003}).

If the X-ray lightcurve was complex, it is not so surprising that the
candidate counterpart did not vary as expected.  For example, in XRN
XTE~J1550-564 the optical ``reflare'' was brighter than the optical
flux during the initial X-ray outburst by more than half a magnitude
\citep{jain2001}.  Another flaring event in this source exhibited a
secondary peak in the optical that was not observed in X-rays at all
\citep{jain2001l}.

Source r3-127 could be a good candidate for such a reflare as the
observed 0.3--7 keV luminosity of the XRN (1.1$\times$10$^{37}$ erg
s$^{-1}$) is only $\sim$10\% of the Eddington luminosity of a neutron
star and only $\sim$1\% of the Eddington luminosity of a typical
stellar-mass black hole.  These numbers suggest that only a small
fraction of the mass of the disk was consumed by the outburst,
allowing the potential for another accretion event that could have
resulted in a high optical flux in the second epoch of ACS data.  All
of these arguments point to our candidate, the most clearly variable
star in the r3-127 error ellipse, as the most likely optical
counterpart to r3-127.

With a strong candidate counterpart, we can predict the orbital period
of r3-127.  There is an empirical relation between the X-ray
luminosity, optical luminosity, and orbital period of Galactic LMXBs
during outburst \citep{vanparadijs1994}.  The relation suggests that
outbursts that are fainter in the optical have smaller accretion disks
due to closer binary separation.  Therefore the fainter the optical
counterpart, the shorter the orbital period of the binary.  

More recently studied LMXBs have also been found to fit this
relation. \citet{williams2005bh1} applied the relation to the
brightest observed X-ray and optical luminosities in a single outburst
for several recently-discovered transient LMXBs.  For example, the
brightest optical and X-ray luminosities observed for 4U~1543-47
during the 1983 outburst were applied to the relation regardless of
the specific timing of the observations. In most cases, the sources
followed the relation.  Furthermore, \citep{williams2005bh4} performed
detailed checks of the application of the relation when applied to
optical and X-ray observations separated by weeks.  For example, the
photometry for A0620-00 from \citet{esin2000} and for XTE~J1550-564
from \citet{jain2001} provided period predictions that were correct
within the (rather substantial) errors.  The X-ray and optical
luminosities of r3-127 therefore provide a rough prediction of the
orbital period of the binary system.
 
The apparent complexity of the X-ray and optical lightcurves of r3-127
adds uncertainty to our application of the \citet{vanparadijs1994}
relation, as we do not have precisely simultaneous X-ray and optical
observations.  On the other hand, the uncertainty introduced by the
non-simultaneous nature of our measurements is small compared to the
dynamic range of the relation.  The relation holds over 8 optical
magnitudes and 3 orders of magnitude in X-ray luminosity, and the
optical variability between observations of the counterpart candidate
is only 1 mag.  Additionally, the intensity of both peaks in X-ray
flux overlapped at the 1$\sigma$ level.  Therefore, even considering
the complexity of the lightcurve and possibility of fast flaring, the
brightest observed optical and X-ray luminosities can provide a
reasonable orbital period prediction.

Applying the \citet{vanparadijs1994} relation requires measurements of
the optical and X-ray luminosity during outburst.  Assuming the same
foreground absorption we used for the X-ray spectral fits, and
applying the conversion of \citet{predehl1995}, the optical extinction
toward r3-127 is $A_B = 0.44$.  In addition, the mean $B-V$ color of
LMXBs in the \citet{liu2001} catalog is -0.09$\pm$0.14.  Correcting
for the extinction, color, and distance to r3-127, the brightest
measured M$_V$ for the counterpart candidate ($B=25.29\pm0.03$), was
M$_V = 0.47\pm0.14$.  The brightest observed 0.3--7 keV luminosity of
r3-127, according to our best-fitting X-ray spectral model, was
L$_X$=1.1$\times$10$^{37}$ erg s$^{-1}$.  These optical and X-ray
luminosities yield an orbital period prediction of
$P=2.3^{+3.7}_{-1.2}$ days including the errors of the relation.  The
final errors change to $P=2.3^{+0.6}_{-0.4}$ days if the errors in the
empirical relation are not taken into account.

Succinctly, the data show that r3-127 is not likely an HMXB, and
therefore it is likely an LMXB.  In addition, the complex X-ray
lightcurve makes the possibility of a complex optical lightcurve more
likely, suggesting the bright variable star could be the optical
counterpart even though it did not fade between observations.
Considering the complex nature of the lightcurve, the highest X-ray
and optical luminosity measurements provide our best prediction for
the orbital period of the system: $P=2.3^{+3.7}_{-1.2}$ days.  This
range is consistent with the periods of many Galactic LMXB transient
systems.

\section{Conclusions}

We have discovered a new bright transient X-ray source in M31 at
R.A.=00:43:09.940 $\pm 0.65''$, Dec.=41:23:32.49 $\pm 0.66''$.  We
have named the source CXOM31~J004309.9+412332 and given it the shorter
name of r3-127.  This source was active for at least 5 months during
2004, but it has not been seen before or since, even in surveys with
limiting fluxes nearly a factor of 100 fainter than our brightest
detection of r3-127.  The observed X-ray lightcurve was double-peaked,
hinting that this source may be of a similar nature to some Galactic
XRNe with complex lightcurves like XTE~J1550-564 or GRO~J1655-40
\citep{mcclintock2004}.

The X-ray spectrum of the source was soft, as is typical for LMXBs.
It was well-fitted by both absorbed power-law and absorbed disk
blackbody models, although the absorbed disk blackbody fits were
somewhat better.  The spectrum appeared marginally softer during the
second half of the 5-month outburst.  The best fit model had an inner
disk temperature of 0.8 keV and an absorption-corrected 0.3--7 keV
peak luminosity of 1.1$\times$10$^{37}$ erg s$^{-1}$ on 04-Oct-2004.

Coordinated {\it HST} ACS observations of the position of r3-127
revealed no stars brighter that M$_B\sim0.1$, confirming r3-127 is not
an HMXB.  The observations also detected one variable star at high
confidence within the error ellipse of the X-ray outburst.  This star
is the strongest candidate counterpart of the X-ray transient source.
The counterpart candidate brightened when our data suggest the X-ray
source was faint, but the complexity of the lightcurve suggests such
optical variations are reasonable.  The highest optical and X-ray
luminosity measurements yield a prediction for the orbital period of
r3-127 of 2.3$^{+3.7}_{-1.2}$ days.  This predicted period range is
large, and it includes the periods of many Galactic LMXB transient
systems.

The location and outburst date of r3-127, along with those of the
other transient sources followed with {\it HST} in 2004 (s1-86, r2-70,
and r2-71; \citealp{williams2005bh1,williams2005bh2,williams2005bh4},
respectively), are consistent with the spatial distribution and rate
of transients measured in \citet{williams2004hrc}.  We continue to
find $\lap$1 new transient source each month, and about half of them
are in the bulge (within $\sim$7$'$ of the nucleus). None of these
recent events has appeared in a cluster or shown a high-mass
secondary, suggesting that the outbursts are not associated with star
formation.

Support for this work was provided by NASA through grant number
GO-9087 from the Space Telescope Science Institute and through grant
number GO-3103X from the {\it Chandra} X-Ray Center.  MRG acknowledges
support from NASA LTSA grant NAG5-10889.  JEM acknowledges support
from NASA ADP grant NNG-05GB31G.

\begin{deluxetable}{ccccccccccc}
\tablecaption{{\it Chandra} ACIS-I observations}
\tableheadfrac{0.01}
\tablehead{
\colhead{{ObsID}} &
\colhead{{Date}} &
\colhead{{R.A. (J2000)}} &
\colhead{{Dec. (J2000)}} &
\colhead{{Roll (deg.)}} &
\colhead{{Exp. (ks)}}
}
\startdata
4680 & 27-Dec-2003 & 00 42 44.4 & 41 16 08.3 & 285.12 & 4.2\\
4682 & 23-May-2004 & 00 42 44.4 & 41 16 08.3 & 79.99 & 3.9\\
4719 & 17-Jul-2004 & 00 42 44.3 & 41 16 08.4 & 116.83 & 4.1\\
4720 & 02-Sep-2004 & 00 42 44.3 & 41 16 08.4 & 144.80 & 4.1\\
4721 & 04-Oct-2004 & 00 42 44.3 & 41 16 08.4 & 180.55 & 4.1\\
4723 & 05-Dec-2004 & 00 42 50.0 & 41 17 15.0 & 269.81 & 4.0\\
\enddata
\label{xobs}
\end{deluxetable}

\begin{deluxetable}{ccccccccccc}
\tablecaption{{\it Chandra} ACIS-I detections of r3-127}
\tableheadfrac{0.01}
\tablehead{
\colhead{{Date}} &
\colhead{{Counts}} &
\colhead{{Flux}\tablenotemark{a}} &
\colhead{{HR1}\tablenotemark{b}} &
\colhead{{HR2}\tablenotemark{c}} &
}
\tablenotetext{a}{The exposure corrected 0.3--10 keV flux in units of 10$^{-5}$ photons cm$^{-2}$ s$^{-1}$}
\tablenotetext{b}{Hardness ratio calculated by taking the ratio of
M-S/M+S, where S is the number of counts from 0.3--1 keV and M is the
number of counts from 1--2 keV.}
\tablenotetext{c}{Hardness ratio
calculated by taking the ratio of H-S/H+S, where S is the number of
counts from 0.3--1 keV and H is the number of counts from 2--7 keV.}
\startdata
23-May-2004 & 60 & 5.7$\pm$0.8 & 0.61$\pm$0.20 & 0.24$\pm$0.27\\ 
17-Jul-2004 & 22 & 2.1$\pm$0.5 & 0.60$\pm$0.28 & 0.21$\pm$0.34\\
02-Sep-2004 & 14 & 1.8$\pm$0.5 & 0.01$\pm$0.40 & -0.20$\pm$0.48\\
04-Oct-2004 & 82 & 7.4$\pm$0.8 & 0.35$\pm$0.14 & -0.14$\pm$0.18\\
05-Dec-2004 & 6 & 0.6$\pm$0.3 & 0.31$\pm$0.43 & -1.1$\pm$1.3\\
\enddata
\label{flux}
\end{deluxetable}

\begin{deluxetable}{ccccccccccc}
\tablecaption{X-ray position determination of r3-127}
\tableheadfrac{0.01}
\tablehead{
\colhead{{ObsID}} &
\colhead{{R.A. (J2000)}} &
\colhead{{$\sigma_{AL}$\tablenotemark{a} ($''$)}} &
\colhead{{$\sigma_{pos}$\tablenotemark{b} ($''$)}} &
\colhead{{$\sigma_{tot}$\tablenotemark{c} ($''$)}} &
\colhead{{Dec. (J2000)}} &
\colhead{{$\sigma_{AL}$ ($''$)}} &
\colhead{{$\sigma_{pos}$ ($''$)}} &
\colhead{{$\sigma_{tot}$ ($''$)}} &
}
\tablenotetext{a}{The alignment error between the X-ray coordinate system and the optical coordinate system from the LGS, determined by {\it ccmap}.}
\tablenotetext{b}{The position error for the source determined by {\it imcentroid}.}
\tablenotetext{c}{The total error for the position, determined by adding the alignment and position errors in quadrature.}
\startdata
4682 & 0:43:10.067 & 0.09 & 1.31 & 1.31 & 41:23:32.73 & 0.12 & 0.95 & 0.96\\
4721 & 0:43:09.899 & 0.14 & 0.74 & 0.75 & 41:23:32.27 & 0.11 & 0.91 & 0.92\\
\hline
Mean & 0:43:09.940 & \nodata & \nodata &  0.65 & 41:23:32.49 & \nodata & \nodata & 0.66\\
\enddata
\label{xpos}
\end{deluxetable}

\begin{deluxetable}{ccccccccccc}
\tablecaption{Fits to the X-ray spectrum of r3-127}
\tableheadfrac{0.01}
\tablehead{
\colhead{{Date}} &
\colhead{{Model}} &
\colhead{{Norm.\tablenotemark{a}}} &
\colhead{{Parameter\tablenotemark{b}}} &
\colhead{{$\chi^2/dof$}} &
\colhead{{$Q$\tablenotemark{c}}} &
\colhead{{L$_X$\tablenotemark{d}}} 
}
\tablenotetext{a}{Normalization parameter for the model.  For the power-law model the units are photons keV$^{-1}$ cm${^2}$ s$^{-1}$ at 1 keV; for the disk blackbody model the parameter is the quantity ((R$_{in}$/km)/(D/10 kpc))$^2$ $\times$ cos($\theta$). Where $\theta$ is the inclination angle of the accretion disk to the
line of sight, $D$ is the distance to the source, and $R_{in}$ is the
radius of the inner edge of the accretion disk.}
\tablenotetext{b}{The final parameter in the model fit.  For power-law
models, this is the photon index of the spectrum.  For disk blackbody
models, this is the temperature ($kT$) in keV of the inner edge of the
accretion disk.}
\tablenotetext{c}{Probability, based on $\chi^2$ statistics, that the observed spectrum is a sample obtained from a source with an intrinsic spectrum equivalent to the model.}
\tablenotetext{c}{The absorption-corrected 0.3--7 keV luminosity of the source in units of 10$^{36}$ erg s$^{-1}$.}
\startdata
23-May-2004 & disk blackbody & 0.010$\pm$0.009 & 0.9$\pm$0.2 & 3.10/4 & 0.54 & 10\\
23-May-2004 & power-law & (3.2$\pm$0.6)$\times$10$^{-5}$ & 1.9$\pm$0.3 & 6.31/4 & 0.18 & 12\\
04-Oct-2004 & disk blackbody & 0.022$\pm$0.016 & 0.8$\pm$0.1 & 4.28/5 & 0.51 & 11\\
04-Oct-2004 & power-law & (4.3$\pm$0.6)$\times$10$^{-5}$ & 2.0$\pm$0.2 & 4.42/5 & 0.49 & 16\\
\enddata
\label{spectab}
\end{deluxetable}

\clearpage
\newpage

\end{document}